# Comparing different models of aftershock rate decay: the role of catalog incompleteness in the first times after main shock

**Barbara Lolli and Paolo Gasperini**

Dipartimento di Fisica

Università di Bologna

Viale Berti-Pichat 8

I-40127 Bologna (Italy),

e-mail: barbara@ibogfs.df.unibo.it, paolo@ibogfs.df.unibo.it

## Abstract

We evaluated the efficiency of various models in describing the time decay of aftershock rate of 47 simple sequences occurred in California (37) from 1933 to 2004 and in Italy (10) from 1976 to 2004. We compared the models by the corrected Akaike Information Criterion (AICc) and the Bayesian Information Criterion (BIC), both based on the log-likelihood function but also including a penalty term that takes into account the number of independent observations and of free parameters of each model. These criteria follow two different approaches (probabilistic and Bayesian respectively) well covering the wide spectra of current views on model comparison. To evaluate the role of catalog incompleteness in the first times after the main shock, we compared the performance of different models by varying the starting time $T_s$ and the minimum magnitude threshold $M_{min}$ for each sequence. We found that Omori-type models including parameter $c$ are preferable to those not including it, only for short $T_s$ and low $M_{min}$ while the latters generally perform better than the formers for $T_s$ longer than a few hours and $M_{min}$ larger than the main shock magnitude $M_m$ minus 3 units. For $T_s$>1 day or $M_{min}$>$M_m$-2.5, only about 15% of the sequences still give a preference to models including $c$. This clearly indicates that a value of parameter $c$ different from zero does not represent a general property of aftershock sequences in California and Italy but it is very likely induced in most cases by catalog incompleteness in the first times after the main shock. We also considered other models of aftershock decay proposed in the literature: the Stretched Exponential Law in two forms (including and not including a time shift) and the band Limited Power Law (LPL). We found that such models perform worse than the Modified Omori Model (MOM) and other Omori-type models for the large majority of sequences, although for LPL, the relatively short duration of the analyzed sequences (one year) might also contribute to its poor performance. Our analysis demonstrates that the the MOM with $c$ kept fixed to 0 represent the better choice for the modeling (and the forecasting) of simple sequence behavior in California and Italy.

## Introduction

As well known since the pioneer work by Omori (1894), the aftershock rate is roughly proportional to the inverse of the time $t$ elapsed after the main shock.

$$\lambda(t) = \frac{K}{t+c} \tag{1}$$

where $K$ is a constant that depends on the total number of aftershocks in the sequence and $c$ a temporal shift. A modification of such law have been later proposed by Utsu (1961) which called it Modified Omori Model (MOM)

$$\lambda(t) = \frac{K}{(t+c)^p} \tag{2}$$

where $p$ and $c$ are free parameters that can be estimated from the data, maximizing the likelihood function of the process (Ogata, 1983). Such power-law behavior is indicative of a physical process generally slower than those typically observed in nature, which are usually described by negative exponentials (Utsu et al., 1995). Although a complete theory of aftershock generation is not available, some simple theoretical hypotheses allow in many cases to deduce laws of this kind (Dieterich, 1986; Yamashita and Knopoff, 1987). A number of successive studies carried on in various part of the world have shown how the MOM satisfactorily describes the time decay of aftershock rate for most sequences excepting those for which the production of secondary aftershocks is particularly significant. In such cases, the use of stochastic branching processes (Hawkes and Adamopulos, 1973; Hawkes and Oakes, 1974; Kagan and Knopoff, 1987; Ogata, 1988) is more appropriate. However in the present paper we restrict our interest to models that can be easily used for the expeditious forecasting of simple sequences not including strong secondary clustering as done for example by Reasenberg and Jones (1989) for California, Eberhart-Phillips (1998) for New Zealand and Lolli and Gasperini (2003) for Italy.

The role of the two free parameters of the MOM is rather different: while $p$ has been introduced by Utsu (1961) to model sequence behaviors significantly deviating from an hyperbolic decay, the inclusion of $c$ was essentially an algebraic expediency to avoid that the rate tends to the infinity for $t$ approaching zero (Gross and Kisslinger, 1994; Narteau et al., 2002). On the other hand, Kagan and Knopoff (1981) justify this divergence at $t$=0 by considering the main shock as the superposition of an infinite number of shocks occurring in an infinitesimal time interval. Nyffenegger and Frohlich (2000) report that limiting the data to an adequate time interval, the Omori model with fixed $c$=0 would adequately describe sequence behavior.

Notwithstanding these doubts on its physical meaning, parameter $c$ is usually assumed different from 0 by most, due to the convenience of eliminating the divergence at $t$=0. However, its inclusion in the rate formulation has also the effect to smooth the decay of the modeled rate with time in the first times after the main shock, particularly for $t<c$. In this time interval the main shock coda and the superposition of many aftershocks often prevent the reliable detection of seismic-phase arrival-times and the measure of phase amplitudes by seismic observatories thus making the location and sizing of all aftershocks quite problematic in the practice. Thus the Maximum Likelihood Estimate (MLE) of $c$ deduced from the data of real sequences might reflect more the incompleteness of the earthquake catalog than the true behavior of the aftershock time-decay (Utsu et al., 1995; Narteau et al., 2002). Vidale et al. (2003; 2004) and Peng and Vidale (2004) note that the number of aftershocks in the first few minutes of an aftershock sequence, observed on high-pass filtered seismograms, is several times higher than the aftershock numbers recorded even in local catalogs.

Moreover, as also discussed by Gasperini and Lolli (2005), the two parameters are strictly correlated and thus the MLE of parameter $c$ might significantly influence the estimate of $p$. This correlation might also result in biased estimates of average parameters to be used as *a-priori* values for the forecasting of future sequences.

The interplay between $p$ and $c$ can be described by considering the following log-log transformation for the time dependent part of eq. (2)

$$\begin{cases} x = \ln t \\ y = \ln\left[(t+c)^{-p}\right] = -p\ln\left[\exp(x)+c\right] \end{cases} \tag{3}$$

the first derivative of $y$ with respect to $x$

$$y' = \frac{dy}{dx} = -p\frac{\exp(x)}{\exp(x)+c} = -p\frac{t}{t+c} \tag{4}$$

describes the slope of the of the MOM log-log curve as a function of time (Fig. 1). We can consider $p_{\text{eff}} = -y'(t)$ as the apparent ("effective") power law exponent of the MOM at time $t$. $p_{\text{eff}}$ is 0 at t=0, it is $1/2p$ at $t=c$ and tends asymptotically to $p$ as $t$ tends to the $\infty$. As well $p_{\text{eff}}$ is constant and coincides with $p$, if $c=0$ ($t\neq0$). In general a high value of $c$ (maybe due to data incompleteness) induces a decrease of $p_{\text{eff}}$ for a relatively long time after the main shock that can be counterbalanced by a corresponding increase of $p$. The effect is instead almost negligible for low $c$ values. The result of this interplay is that $p$ and $c$ are positively and almost linearly correlated, as Gasperini and Lolli (2005) evidenced through a statistical analysis of independent parameter estimates of sequences from Italy (Lolli and Gasperini, 2003) and New Zealand (Eberhart-Phillips, 1998). Gasperini and Lolli (2005) also showed how in some cases the MLEs of $p$ and $c$ are only barely constrained by the data, along an almost straight trajectory of the parameter space.

Other functional forms have been proposed in the past to describe the decay of aftershock rate with time. On the basis of physical considerations, originally made by Shlesinger and Montroll (1984), for a physical problem -the dielectric relaxation- that presents many analogies with aftershock decay, Kisslinger (1993) proposed the stretched exponential law (STREXP)

$$\lambda(t) = qN^*(0)\frac{1}{t}\left(\frac{t}{t_0}\right)^q \exp\left[-\left(\frac{t}{t_0}\right)^q\right] \tag{5}$$

where $N^*(0)$ is the number of "potential aftershock sites" at $t$=0 (corresponding to the total number of aftershock for $t\to\infty$) while $t_0$ and $q$ are characteristic parameters depending on the physical properties of the area. For $q$=1 the model becomes the classical exponential decay (Debye relaxation) which has also been suggested as a model for aftershock decay (Mogi, 1962) while for $q$=0 it corresponds to the simple Omori's model.

The STREXP model has been modified by Gross and Kisslinger (1994) to include, analogously to the MOM, a time shift $d$

$$\lambda(t) = qN^*(0)\exp\left[\left(\frac{d}{t_0}\right)^q\right]\frac{1}{t+d}\left(\frac{t+d}{t_0}\right)^q \exp\left[-\left(\frac{t+d}{t_0}\right)^q\right] \tag{6}$$

Gross and Kisslinger (1994) showed that, the STREXP and its modification might fit the data of some Californian sequences better than the MOM.

More recently, Narteau et al (2002, 2003) proposed a "band limited" power law (LPL) restricted to the interval between two characteristic rates, above and below which the decay with time is respectively linear and exponential

$$\lambda(t) = \frac{A\left[\gamma(q,\lambda_a t) - \gamma(q,\lambda_b t)\right]}{t^q} \tag{7}$$

where

$$\gamma(\rho,x) = \int_0^x \tau^{\rho-1}\exp(-\tau)d\tau \tag{8}$$

is the incomplete Gamma function, and $A$, $q$, $\lambda_a$ and $\lambda_b$ are free parameters of the model. In particular $A$ has a role similar to parameter $K$ of the Omori's model and is chosen so that the time integral of the rate from the beginning to the end of the sequence equates the total number of shocks; $q$ is the exponent of the power law decay (similar to parameter $p$ of Omori's models) while $\lambda_b >> \lambda_a$ are the values of the rate at which the mode of decay changes from linear to power-law and from power-law to exponential respectively. Assuming a given threshold (*i.e.* 0.8, 0.9, 0.99) for the rate divergence $\zeta$ between LPL and a pure power-law, Narteau et al. (2002, 2003) determine two transition times $t_b^\zeta < t_a^\zeta$ separating the three stages of the process. This model has two very interesting properties: the rate integral is convergent even if $q$<1 and the rate is finite at $t$=0.

In this work we will analyze a set of aftershock sequences occurred in Italy and California to investigate on the real significance of parameter $c$ in the MOM and particularly on the effects of catalog incompleteness in the first times after the main shock on *c* estimates. To do that, we will compare the performance of Omori-type models including parameter *c* and not including it. We will also consider in the comparison the alternative decay models (eq. 5, 6 and 7) described above.

Catalog incompleteness in the first times after the main shock might be due to several factors (see Kagan, 2004, for a comprehensive discussion) ranging from the overlapping of seismic records that makes difficult the identification and location of many shocks, to the changes in the seismographic network (some stations might be damaged by the main shock). Moreover, as the number of shocks to be located and sized might be several order of magnitude larger than in quiet periods, the seismic network personnel might be unable to sustain the workload and thus the network management might take the administrative decision to increase the magnitude threshold above the standard one. In all these cases catalog incompleteness should be a decreasing function of time after the main shocks (as the risk of superposition and neglect decreases with the decreasing rate of aftershocks) as well as of the minimum magnitude threshold (as stronger aftershocks are more likely to be recognized and analyzed well by network operators). We will thus estimate the parameters and compare the goodness of fit of various models by varying the starting time of the modeled portion of the sequences and the minimum magnitude of aftershocks considered for the modeling. In any case we will restrict our analysis to the time interval following the first detected aftershock above minimum magnitude thus we will be unable to make inferences on the behavior of the process at times comparable with rupture time of the main shock where the point model has to somehow break down.

## Catalog data and sequence detection

For California we simply used the revised catalog from 1932 to 2004 available from the Southern California Earthquake Center (SCEC) site (http://www.scecdc.scec.org/). For Italy, we had instead to merge several catalogs of Italian instrumental earthquakes covering the time interval from 1960 to 2004. From 1981 to 1996, the data come from the *Catalogo strumentale dei terremoti Italiani dal 1981 a 1996* Version 1.1 (CSTI Working Group, 2001; 2004), which is based on an integration of the seismic phase data of the *Rete Sismica Nazionale Centralizzata* (RSNC) of the *Istituto Nazionale di Geofisica e Vulcanologia* (INGV) and of some regional networks operating on the Italian territory. In CSTI, all earthquakes have been relocated with uniform methods and the magnitudes revalued from original amplitudes and coda-durations, according to Gasperini (2002), by the calibration with real and synthetic Wood-Anderson magnitudes. From 1997 to 2002 we used the *Catalogo della Sismicità Italiana* 1.0 (Castello et al, 2005) which also is based on the relocation of an integrated dataset of RSNC and regional networks with revised magnitudes. For the period from 1960 to 1980, we used the catalog of the *Progetto Finalizzato Geodinamica* (Postpischl, 1985). For this dataset, we corrected, according to Lolli and Gasperini (2003), the original instrumental magnitudes on the basis of an empirical comparison of the observed rates with respect to the CSTI recalibrated catalog from1981 to 1996. Finally from 2003 to 2004 data are taken from the instrumental bulletin of the INGV available from site http://www.ingv.it/~roma/reti/rms/bollettino. The integrated catalog from 1960 to 2004 in CIT format can be downloaded from the authors' anonymous ftp site: ftp://ibogfs.df.unibo.it/lolli/aft2005/.

To build the sequences, we only considered the shocks shallower than 40 km. We define the space-time influence zone of any shock as a spatial circular area centered in the epicenter and a time window starting at the shock origin time. We consider as main shocks (originating sequences) all earthquakes with magnitude not lower than 5.0 and not included in the influence zone of a larger shock. The time window is fixed to one year for all main shocks while the radius $R$ varies as a function of magnitude as $\mathrm{Log}_{10}(R)=0.1238\mathrm{M}+0.983$ (corresponding to Table 1 of Gardner and Knopoff, 1974). Considering the need to analyze subsets of the data with varying starting times and minimum magnitudes we have selected only the sequences (listed in Table 1) including at least 100 shocks with magnitude not lower than main shock magnitude $M_m$ minus 3.5 (93 for the large Southern Italy earthquake of November 23rd 1980). Moreover, since all the decay models described above are not suitable to model complex sequences with evident secondary clustering due to strong aftershocks, we excluded from computations all sequences (marked with an asterisk in Table 1) including at least an aftershock with magnitude larger than main shock magnitude minus 0.6 (one half of the average magnitude difference between the main shocks and the largest aftershocks, according to Bath (1965) law). In this way we selected in all 37 sequences (over 61 detected) for California and 10 sequences (over 15 detected) for Italy. The corresponding files in CIT format are also available from authors' anonymous ftp site.

## Model fitting and comparison

We used the maximum likelihood procedure (Ogata, 1983) to estimate the best fitting parameters for the different models of aftershock decay rate. This approach consists of approximating the difference between the $N$ aftershocks occurring at times $t_i$, ($i$=1,$N$), during the time interval [$T_s$, $T_e$] and an intensity function $\lambda(t)$ defined from a non-stationary Poisson process (Narteau et al., 2002). The log-likelihood function of such a process for a parameter set $\theta$ is given by (Ogata, 1983)

$$l(\theta) = \sum_{i=1}^{N} \ln\left[\lambda(\theta; t_i)\right] - \int_{T_s}^{T_e} \lambda(\theta; t) dt \tag{9}$$

The parameter set $\hat{\theta}$ which maximize eq. (9) for the aftershock times $t_i$ is the MLE of model parameters for the given sequence. We maximized Equation (9) in two steps: we first explored a regular grid of parameter values covering reasonable ranges of variation for each of them. Successively the approximate maximum was refined using the Fortran routine BCONG/DBCONG of the IMSL Math library (Visual Numerics, 1997). This code uses a quasi-Newton method (Dennis and Schnabel, 1983) and an active set strategy (Gil and Murray, 1976) to solve optimization problems subject to simple bounds.

For each sequence, we considered the rate equations for the original Omori model (eq. 1) and the MOM (eq. 2) as well as for two further models obtained from the previous ones by removing $c$

$$\lambda(t) = \frac{K}{t} \tag{10}$$

$$\lambda(t) = \frac{K}{t^p} \tag{11}$$

The MLE for the four models are computed for increasing starting times $T_s$ with logarithmic steps going from 0.001 to 1.79 days and varying the minimum magnitude $M_{min}$ with linear steps of 0.1 magnitude units from main shock magnitude $M_m$ minus 3.5 and $M_m$-2.5.

The comparison among different empirical models must take into account not only the goodness of fit with data but also the number of free parameters of each model. In fact, as the data are noisy, there is the possibility that complex models might overfit random features of data not representing true physical properties of the investigated process. Thus the “best model” is not necessarily the one showing the absolute maximum likelihood with data but a simpler model could be preferable if the contribution of some of its parameters to the likelihood improvement is negligible. This concept is well described by the General Information Criterion (GIC) (Leonard and Hsu, 1999) where a “penalty” term for each parameter is subtracted from the maximized log-likelihood function

$$GIC = \ln l(\hat{\theta}) - \frac{\alpha}{2} k = L(\hat{\theta}) - \frac{\alpha}{2} k \tag{12}$$

where $\hat{\theta}$ is the set of parameter values maximizing the log-likelihood function $L(\theta)$, $k$ is the number of free parameters and $\alpha$ the penalty assigned to each parameter. With an appropriate choice of $\alpha$, GIC can be used to find the model which fit data at best using the minimum possible number of free parameters, among a set of alternatives: the better the model the larger the GIC. The choice of the penalty is not univocal in the literature. In the initial proposition of the method, Akaike (1974) deduced, that at the first order, $\alpha$=2 is the correct asymptotic bias correction term which minimize the Kullback-Leibler information or distance (Burnham and Anderson, 1998). This choice (apart for a multiplicative factor -2) leads to the well-known Akaike information criterion (AIC) already used by several authors (*i.e.* Ogata, 1983; Gross and Kisslinger, 1994; Narteau et al., 2003) to compare the performance of aftershock decay models

$$AIC = \max L\left(\hat{\theta}\right) - k \tag{13}$$

Sugiura (1978) showed that AIC might perform poorly if there are too many parameters in relation to the size of the sample. He derived a second-order variant of AIC that he called c-AIC. Hurvich and Tsai (1989) further studied this small-sample bias adjustment, which led to a criterion, assuming $\alpha=2n/(n-k-1)$, that is called $AIC_c$

$$AIC_c = \max L\left(\hat{\theta}\right) - k - \frac{k(k+1)}{n-k-1} = AIC - \frac{k(k+1)}{n-k-1} \tag{14}$$

where $n$ is the number of data in the sample. The use of $AIC_c$ is always preferable with respect to AIC if the ratio between the number of data and the number of parameters $n/k$ (as in most of our sequences) is lower than 40 (Burnham and Anderson, 1998).

Under slightly different assumptions, Schwarz (1978) showed instead, that the choice $\alpha=\ln n$ is asymptotically optimal since, for $n\rightarrow\infty$, the probability that the best model have the higher BIC tend to the unity. Furthermore Draper (1995) noted that the addition of a term $-\ln(2\pi)$, neglected by Schwarz (1978), may improve the accuracy. The estimators resulting from both these choices are usually referred in the literature as Bayesian Information Criterion (BIC) (*i.e.* Burnham and Anderson, 1998; Leonard and Hsu, 1999; Main et al., 1999). For sake of clarity we will indicate them as Schwarz information criterion (SIC) and BIC respectively

$$SIC = \max L\left(\hat{\theta}\right) - \frac{k}{2}\ln n \tag{15}$$

$$BIC = \max L\left(\hat{\theta}\right) - \frac{k}{2}\ln\frac{n}{2\pi} \tag{16}$$

The four estimators (AIC, $AIC_c$, SIC and BIC) work similarly in many cases but might give slightly different results, depending on the sample size and on the number of free parameters. In general AIC and $AIC_c$ tend to prefer simpler models than SIC and BIC for smaller sample sizes, and more complex models for larger samples. In particular AIC is equivalent to SIC for $n\approx7$ while BIC stays about in the middle among the two, as it is equivalent to AIC for $n \approx 46$. For $k$ ranging from 1 to 5, $AIC_c$ is equivalent to SIC for $n$ ranging between 11 and 19 and to BIC for $n$ between 50 and 58.

For simplicity, we will consider, in the following only $AIC_c$ and BIC as they adequately represent the two alternative approaches (probabilistic and Bayesian) to information theory in model evaluation. We will use both of them to assess, for each sequence, the best among the four Omori-type models (eq. 1, 2, 10 and 11) as well as among such best model and the alternative ones: STREXPs and LPL (eq. 5, 6 and 7). For sake of a comparison with previous papers (Gross and Kisslinger, 1994; Narteau et al., 2003) we will also compare the performance of the STREXPs and the LPL with the MOM (eq. 2). The numbers of free parameter ($k$) to be counted by $AIC_c$ (eq. 14) and BIC (eq. 16) are 1 for the model of eq. (10), 2 for the original Omori model (1) and the model of eq. (11), 3 for the MOM (eq. 2) and the STREXP (eq. 5), 4 for the modified STREXP (eq. 6) and the LPL (eq. 7).

## Results and Discussion

In Table 1 we report the MLE of MOM parameters, for the selected Californian and Italian sequences, assuming as $T_s$ the time of the first aftershock above the minimum magnitude threshold $M_{min}=M_m$-3.5. Parameter standard deviations are computed from the diagonal elements of the inverse of the Fisher information matrix (see *i.e.* Guo and Ogata, 1997). We can note how for 14 Californian and 2 Italian sequences respectively, the best $c$ estimate (maximizing the likelihood) is exactly 0 (the standard deviations are not computed as such value is at the edge of the definition interval). This indicates that for these sequences, eq. (11) is preferable with respect to MOM (eq. 2) independently on the incompleteness of the catalog in the first times after the main shock.

In the following we will count the number of times each model is found to be the best among the alternatives, on the basis of AICc and BIC criteria, and we will plot the behavior of these counts as a function of starting time $T_s$ and minimum magnitude $M_{min}$. In Fig. 2, the comparison concerns the four Omori-type models (eq. 1, 2, 10, 11) as a function of the starting time $T_s$ for sequences of California and $M_{min}=M_m$-3.5. For $T_s$<0.01 days (about 15 minutes) the models including $c$ (eq. 1 and 2) prevails for the most of the analyzed sequences. Particularly the MOM (green) and the model with $p$ free and $c$ fixed to 0 (black) are almost in balance but among the models with fixed $p$=1, the one with $c$ free (red) clearly prevails with respect to the one with $c$=0 (blue). Starting from about 0.01 days after the main shock, we can note a clear decrease of choices in favor of models including $c$ and a corresponding increase of the preferences for models without $c$ (eq. 10 and 11). After 1 day the counts become 34 versus 3 in favors of the latter ones. Even the simple hyperbolic law (blue) appears to be preferable with respect to both models with *c* free. A similar behavior is shown in Fig. 3 for Italy. For this dataset the preferences for models with *c* free are more than those for models with *c* fixed to 0, up to $T_s$≈0.1-0.3 days (2.5-7 hours) after the main shock. After 1 day the balance becomes about 8 to 2 in favor of models without *c* and at the maximum $T_s$=1.79 days the mostly preferred model is the simple hyperbolic one in 4 cases over 10. For both datasets, AICc and BIC show similar behaviors with maximum differences of 5 counts for California and 2 for Italy. Thus for sake of brevity we will show only BIC in the following figures.

Figure 4 shows the results for both datasets put together. The scores after 1 day indicate the preference of models without *c* for 42 sequences versus 5. In all cases shown the counts for the MOM (the most complicate model) tend almost to 0 for long starting times. Note that, for all sequences and thresholds, the MOM, which comprises within it the other simpler models, reaches anyhow the maximum log-likelihood among all models (including some ties with model of eq. 11 when the estimated *c* is exactly 0). The better performance of the simpler models is, in fact, due to the penalty term assigned by $AIC_c$ and BIC to the additional parameters of the MOM.

These results seem to indicate that in most cases a value of *c* different from 0 is strongly related to the incompleteness in the first hours after the main shock. Particularly, the Italian catalog appears to be slightly more affected by incompleteness with respect to the Californian one.

In Figure 5 the best Omori-type model (a) and the MOM (b) are compared (using BIC criterion) with other alternative models (eq. 5, 6 and 7). In this case the preference for the former ones is clear and only slightly dependent on starting time $T_s$. In fact, although in several cases the log-likelihood of alternative models (and particularly of LPL) is higher than the one of the best Omori-type model or of the MOM, the improvement of the fit does not justify the increase of free parameters. We must note however that the low performance (in terms of $AIC_c$) of LPL can be justified by the method we have adopted to select sequences. In fact, such model might be preferable with respect to the MOM when the observing time interval is long enough to include the transition from power-law to exponential decay and this might not be the case due to the limit of one year we imposed to sequence duration. On the other hand our analysis seems to indicate, in agreement with previous findings, that both the STREXP models are only rarely preferable to MOM.

In Fig. 6, for the joint set of Californian and Italian sequences, we focus on the comparison between the two Omori type models having both *p* free to vary and *c* free (the MOM, eq. 2, solid curve) and fixed to 0 (eq. 11, dotted) respectively. The crossover starting time is about at 0.006 days (8 minutes) and, for $T_s$ larger than a few hours after the main shock the scores became about 8 times to 1 in favor of the model without *c*.

To better investigate on the nature of this preference for the model with $c$ fixed to 0 we performed additional comparisons varying the minimum magnitude $M_{min}$ threshold from $M_m$-3.5 to $M_m$-2.5. In Fig. 7 and 8 we show the results of this comparison for sequences from California and Italy respectively, by assuming as $T_s$ the time of the first aftershock. For both datasets the preferences are almost in balance at the lower threshold $M_m$-3.5 with a slight prevalence of the MOM, but the model with fixed $c$=0 starts to progressively increase its counts for increasing thresholds and clearly performs better than the MOM with $c$ free for $M_{min}>M_m-3$. For the California dataset we have the crossover at $M_m$-3.4 and a clear divergence of the two curves starting from $M_m$-3.0 while for Italy, after the initial prevalence of the MOM, there is a tie up to $M_m$-3.0 and a clear divergence starting from $M_{min}=M_m-2.7$. Apart from small differences among the two datasets (the Italian catalog appears again less complete than the Californian one) these computations clearly confirms that for the great majority of the sequences the incompleteness of the catalog in the first times after main shock is likely to be the cause of $c$ values different from 0.

In order to show the effect of time and magnitude thresholds on the parameters of decay models, we reported, for all the analyzed sequences, the plot of the behavior of $p$ parameter for eq. (2) and (11) as well of $c$ for eq. (2) as a function of starting time $T_s$ and of the difference between minimum and main shock magnitudes $M_{min}-M_m$ in figure 9 and 10 respectively. Both in Fig. 9a and 9b we can observe a general increase of the scatter of parameter $p$ among different sequences for increasing starting time although the $p$ values as well as the scatter are generally lower for model of eq. (2) (Fig. 9a) with respect to the MOM (Fig. 9b). This difference is a clear empirical confirmation of the effect of the correlation between the two parameters of the MOM described above: a $c$ value different from 0 implies a larger MLE of $p$. As well, the interplay between the two parameters increases the dispersion of $p$ estimates. In both plots the thick dashed lines show the behaviors of the arithmetic average of $p$ values as a function of starting time. In both cases the averages appear rather stable but, for model of eq. (2), we can see a initial increase from 0.74 at short $T_s$ to about 0.76 for $T_s$=0.1 days followed by a decrease down to 0.71 for the maximum starting time of 1.8 days while for the MOM the trend is almost monotonic and ranges from about 0.82 at $T_s$=0.001 days to 0.89 at $T_s$=1.8 days. On the other hand the $c$ values show in many cases (when they are different from 0) an almost constant behavior, then a drop to low values (or to 0) which sometimes is followed by an oscillation, indicating an instability of such parameter for large starting times after main shock. In this case the thick dashed line indicates the geometric average of non-null $c$ estimates at every $T_s$. Considering the direct correlation between the two parameters of the MOM, the increasing trend of the $c$ average might explain the absence, for the MOM, of the decrease of the $p$ average observed for model of eq. (2).

In fig. 10a we can note for model of eq. (2) a general increasing trend of $p$ for increasing minimum magnitude while for the MOM (Fig. 10b) the same increasing behavior is somehow masked, for some sequences, by the interplay with parameter $c$ (Fig. 10c) that is generally decreasing with a drop to very low (or null) $c$ values starting from different thresholds depending on the sequence. For the MOM the increase of the minimum magnitudes appears to slightly reduce the scatter of parameter estimates among different sequences. The average values show for eq. (2) a significant increase of $p$ from 0.74 to 0.82 when going from $M_{min}= M_m$-3.5 to $M_{min}= M_m$-2.5 and a smaller increase (from 0.82 to 0.84) for the MOM, while the geometric average of non-null $c$ values decrease from about $4x10^{-2}$ to $5x10^{-3}$.

**Conclusions**

Our analysis showed that parameter $c$ of the Modified Omori Model (MOM) is not necessary to fit at best most of the 47 simple aftershock sequences we have studied, occurred in California and Italy from the beginning of the instrumental era to 2004. Although a minority of sequences (about 10-15%) actually requires the presence of $c$, we can confidently assert that such parameter does not represent a general feature of the aftershock rate decay but it is mainly an artifice induced by catalog incompleteness in the first times after the main shock. Both models not including $c$ (eq. 10 and 11) have shown to be preferable with respect to the corresponding models including $c$ (eq. 1 and 2), if the first hours after the main shock, when the catalog is likely to be incomplete, are excluded by the analysis. It is interesting to note how the very primitive hyperbolic model (with $p$=1 and $c$=0) might actually well represent the behavior of many sequences as it is found to be preferable among all others for more that about 1/3 of the Italian sequences (1/5 in California) if the modeling interval starts about 7-8 hours after the main shock.

On the other hand, if the first hours after the main shock are excluded from the analysis and/or the minimum magnitude threshold of shocks included in modeling is taken larger than main shock magnitude minus 3 units, the rate decay model of eq. (11) with variable $p$ and $c$ fixed to 0 is preferable in about 40 cases over 47 with respect to the MOM. In our opinion, such model represents the best choice for modeling the behavior of simple sequences even because its use avoids the danger of a biased estimate of parameter $p$ due to the interplay with parameter $c$ (see Gasperini and Lolli, 2005). Even the problem of the divergence at $t$=0 is not effective in the practice as the most reliable estimates are made using $T_s \neq 0$. Another possible alternative for the modeling of sequence behavior could be the adoption of a procedure derived from the one we have followed in our analysis that is i) excluding from computation the first 6-10 hours after the main shock and/or the shocks with magnitude lower than $M_m$-3 and ii) choosing, for each sequence, the model that perform better, on the basis of $AIC_c$ or BIC. With both alternatives the reliability of the estimated parameters as well as the forecasting ability of the a-priori average model should improve due to the reduction of the danger of overfitting and of the interplay between the two parameters of the MOM.

The behavior of $p$ values with varying starting times and minimum magnitude appears quite stable (the average over all sequences varies from about 0.82 to 0.89 for the MOM and from 0.74 to 0.79 for the model of eq. (2)) although the scatter among different sequences increases for increasing starting time and slightly decreases for increasing minimum magnitude. For the MOM, the comparison with the behavior of $c$ (which is increasing for increasing starting time and decreasing for increasing minimum magnitude) confirms the existence of a strict correlation between the two parameters that, at least in some cases, might be the cause of a biased estimate of $p$.

We also verified that the Stretched Exponential model (Gross and Kisslinger, 1994) does not represent a valid alternative to the Omori-type models as its performance is definitely poorer. On the contrary the same conclusion cannot be drawn, for the band Limited Power Law (LPL) model (Narteau et al., 2003), as the chosen length of the time interval (1 year) could not include the time when the transition from the power-law to the negative exponential actually occurs. We believe that further efforts making use of the LPL could confirm the effectiveness of such model, particularly in the long times portion of the sequences. Our results regarding the short times, showing that the deviation from the power-law is likely to be due to catalog incompleteness in most cases, would suggest a modification of LPL where the short time transition from linear decay to power-law is removed and only the long time transition from power-law to exponential is considered. This can be easily implemented by substituting the first incomplete Gamma function in eq. (7) with the complete one $\Gamma(q)$:

$$\lambda(t) = \frac{A\left[\Gamma(q) - \gamma(q, \lambda_b t)\right]}{t^q} \tag{17}$$

This modified model will be the subject of a future paper currently in preparation.

**Acknowledgements**

We thank Yan Kagan and an anonymous referee for their thoughtful comments and suggestions that helped much to improve the paper. This work was supported by the *Istituto Nazionale di Geofisica e Vulcanologia* (Contracts: GNDT 2000/2004) and by the *Ministero dell'Istruzione, dell'Università e della Ricerca* (COFIN 2002).

**Table captions**

Table 1 – List of analyzed aftershock sequences for California (cal*xx*) and Italy (ita*xx*). $M_m$ is the main shock magnitude, $M_{min}$ the minimum magnitude of aftershocks used in computations with varying starting time, $N$ the total number of shocks above $M_{min}$, *Dur* the effective duration (in days) of the aftershock sequence (corresponding to the time difference between main shock and the last aftershock), $p$, $c$, and $K$ the values of MOM parameters estimated using the time of first aftershock as starting time $T_s$, *erp*, *erc*, and *erK* the corresponding standard deviations. Values in boldface indicate that $c$ parameter is exactly 0. In these cases the standard deviations cannot be computed from the Fisher information matrix as the parameter value is at the edge of definition interval.

## Table 1

| *Seq* | *Year* | *Mo* | *Da* | *Lat* | *Lon* | $M_m$ | $M_{min}$ | *N* | *Dur* | *p* | *erp* | *c* | *erc* | *K* | *erK* |
|---|---|---|---|---|---|---|---|---|---|---|---|---|---|---|---|
| cal01 | 1933 | 3 | 11 | 33.638 | -117.973 | 6.4 | 2.9 | 275 | 357 | 1.01 | 0.03 | 0.05 | 0.02 | 31.90 | 3.29 |
| cal02 * | 1946 | 3 | 15 | 35.742 | -117.912 | 5.4 | 2.2 | 153 | 364 | - | - | - | - | - | - |
| cal03 | 1946 | 3 | 15 | 35.713 | -117.897 | 6.3 | 2.8 | 129 | 364 | 0.93 | 0.11 | 2.57 | 1.88 | 20.29 | 10.57 |
| cal04 * | 1947 | 7 | 24 | 34.023 | -116.477 | 5.3 | 2.3 | 115 | 342 | - | - | - | - | - | - |
| cal05 * | 1948 | 12 | 4 | 33.996 | -116.233 | 6.0 | 2.5 | 291 | 361 | - | - | - | - | - | - |
| cal06 | 1952 | 7 | 21 | 34.981 | -119.017 | 7.5 | 4.0 | 203 | 334 | 1.12 | 0.07 | 0.19 | 0.11 | 33.61 | 7.36 |
| cal07 * | 1954 | 1 | 12 | 34.993 | -119.061 | 5.4 | 2.4 | 111 | 350 | - | - | - | - | - | - |
| cal08 | 1954 | 3 | 19 | 33.286 | -116.069 | 6.4 | 2.9 | 125 | 309 | 1.09 | 0.04 | 0.01 | 0.00 | 12.02 | 1.34 |
| cal09 | 1968 | 4 | 9 | 33.167 | -116.087 | 6.6 | 3.1 | 107 | 354 | 1.03 | 0.06 | 0.04 | 0.03 | 12.39 | 2.18 |
| cal10 | 1971 | 2 | 9 | 34.416 | -118.370 | 6.6 | 3.1 | 281 | 363 | 1.04 | 0.02 | 0.00 | 0.00 | 23.26 | 1.59 |
| cal11 | 1975 | 6 | 1 | 34.512 | -116.488 | 5.3 | 1.8 | 291 | 360 | 0.64 | 0.03 | **0.00** | - | 12.67 | 1.32 |
| cal12 * | 1976 | 1 | 10 | 32.128 | -115.447 | 5.0 | 1.5 | 206 | 357 | - | - | - | - | - | - |
| cal13 | 1976 | 11 | 4 | 33.115 | -115.616 | 5.1 | 1.6 | 310 | 360 | 0.52 | 0.03 | **0.00** | - | 8.81 | 1.17 |
| cal14 | 1978 | 5 | 5 | 32.210 | -115.303 | 5.2 | 1.7 | 254 | 357 | 0.40 | 0.04 | **0.00** | - | 4.53 | 0.81 |
| cal15 | 1978 | 8 | 13 | 34.347 | -119.696 | 5.1 | 1.6 | 379 | 358 | 0.94 | 0.02 | 0.01 | 0.01 | 35.80 | 2.82 |
| cal16 * | 1978 | 10 | 4 | 37.513 | -118.683 | 5.8 | 2.3 | 179 | 363 | - | - | - | - | - | - |
| cal17 | 1979 | 1 | 1 | 33.943 | -118.681 | 5.2 | 1.7 | 270 | 350 | 0.92 | 0.03 | 0.01 | 0.01 | 24.10 | 2.12 |
| cal18 * | 1979 | 3 | 15 | 34.327 | -116.445 | 5.3 | 1.8 | 727 | 360 | - | - | - | - | - | - |
| cal19 * | 1979 | 10 | 15 | 32.613 | -115.318 | 6.4 | 2.9 | 434 | 269 | - | - | - | - | - | - |
| cal20 | 1980 | 2 | 25 | 33.500 | -116.513 | 5.5 | 2.0 | 204 | 362 | 0.62 | 0.03 | **0.00** | - | 8.49 | 1.09 |
| cal21 * | 1980 | 5 | 25 | 37.607 | -118.821 | 6.4 | 2.9 | 952 | 220 | - | - | - | - | - | - |
| cal22 * | 1980 | 5 | 25 | 37.555 | -118.790 | 6.5 | 3.0 | 756 | 220 | - | - | - | - | - | - |
| cal23 * | 1980 | 9 | 7 | 38.038 | -118.267 | 5.0 | 1.8 | 172 | 115 | - | - | - | - | - | - |
| cal24 | 1980 | 9 | 7 | 38.050 | -118.333 | 5.4 | 1.9 | 160 | 115 | 0.60 | 0.05 | 0.01 | 0.04 | 9.87 | 1.74 |
| cal25 | 1981 | 4 | 26 | 33.096 | -115.624 | 5.8 | 2.3 | 339 | 364 | 1.02 | 0.03 | 0.02 | 0.01 | 37.77 | 3.11 |
| cal26 | 1981 | 9 | 4 | 33.557 | -119.120 | 5.4 | 1.9 | 317 | 359 | 0.69 | 0.02 | **0.00** | - | 16.23 | 1.43 |
| cal27 | 1982 | 10 | 1 | 35.743 | -117.756 | 5.1 | 1.6 | 1546 | 363 | 0.67 | 0.01 | 0.01 | 0.01 | 75.49 | 4.01 |
| cal28 * | 1982 | 10 | 25 | 36.291 | -120.404 | 5.6 | 2.1 | 418 | 364 | - | - | - | - | - | - |
| cal29 * | 1983 | 7 | 22 | 36.258 | -120.384 | 5.8 | 2.3 | 585 | 364 | - | - | - | - | - | - |
| cal30 | 1985 | 8 | 4 | 36.151 | -120.049 | 5.7 | 2.2 | 364 | 362 | 0.84 | 0.02 | 0.02 | 0.01 | 28.81 | 2.53 |
| cal31 | 1986 | 7 | 8 | 33.999 | -116.608 | 5.7 | 2.2 | 1329 | 364 | 0.86 | 0.01 | 0.03 | 0.01 | 112.76 | 5.71 |
| cal32 | 1986 | 7 | 13 | 32.971 | -117.874 | 5.5 | 2.0 | 1649 | 364 | 0.70 | 0.01 | 0.02 | 0.01 | 88.11 | 5.14 |
| cal33 | 1987 | 2 | 7 | 32.388 | -115.305 | 5.4 | 1.9 | 295 | 362 | 0.73 | 0.02 | **0.00** | - | 17.04 | 1.44 |
| cal34 * | 1987 | 10 | 1 | 34.061 | -118.079 | 5.9 | 2.4 | 148 | 348 | - | - | - | - | - | - |
| cal35 * | 1987 | 11 | 24 | 33.090 | -115.792 | 6.2 | 2.7 | 555 | 364 | - | - | - | - | - | - |
| cal36 | 1987 | 11 | 24 | 33.015 | -115.852 | 6.6 | 3.1 | 169 | 364 | 1.11 | 0.04 | 0.03 | 0.01 | 19.59 | 2.17 |
| cal37 | 1988 | 6 | 10 | 34.943 | -118.743 | 5.4 | 1.9 | 153 | 362 | 0.44 | 0.05 | **0.00** | - | 3.10 | 0.66 |
| cal38 * | 1988 | 12 | 3 | 34.151 | -118.130 | 5.0 | 1.5 | 191 | 363 | - | - | - | - | - | - |
| cal39 * | 1988 | 12 | 16 | 33.979 | -116.681 | 5.0 | 1.5 | 612 | 363 | - | - | - | - | - | - |
| cal40 | 1989 | 1 | 19 | 33.919 | -118.627 | 5.0 | 1.5 | 359 | 364 | 0.75 | 0.02 | **0.00** | - | 21.78 | 1.57 |
| cal41 | 1990 | 2 | 28 | 34.144 | -117.697 | 5.5 | 2.0 | 619 | 363 | 0.86 | 0.02 | 0.01 | 0.01 | 49.44 | 3.16 |
| cal42 * | 1992 | 4 | 23 | 33.960 | -116.317 | 6.1 | 2.6 | 2699 | 362 | - | - | - | - | - | - |
| cal43 | 1992 | 6 | 28 | 34.200 | -116.437 | 7.3 | 3.8 | 230 | 359 | 1.07 | 0.04 | 0.07 | 0.03 | 29.77 | 3.74 |
| cal44 * | 1993 | 8 | 21 | 34.029 | -116.321 | 5.0 | 1.5 | 1902 | 364 | - | - | - | - | - | - |
| cal45 | 1994 | 1 | 17 | 34.213 | -118.537 | 6.7 | 3.2 | 322 | 346 | 1.20 | 0.04 | 0.09 | 0.03 | 49.10 | 5.67 |
| cal46 | 1994 | 6 | 16 | 34.268 | -116.402 | 5.0 | 1.5 | 2083 | 364 | 0.12 | 0.02 | **0.00** | - | 10.26 | 0.97 |
| cal47 | 1995 | 6 | 26 | 34.394 | -118.669 | 5.0 | 1.5 | 502 | 364 | 0.22 | 0.03 | **0.00** | - | 3.98 | 0.66 |
| cal48 * | 1995 | 8 | 17 | 35.776 | -117.662 | 5.4 | 1.9 | 1374 | 362 | - | - | - | - | - | - |
| cal49 * | 1995 | 9 | 20 | 35.761 | -117.638 | 5.8 | 2.3 | 444 | 364 | - | - | - | - | - | - |
| cal50 | 1996 | 11 | 27 | 36.075 | -117.650 | 5.3 | 1.8 | 1089 | 362 | 0.79 | 0.02 | 0.11 | 0.04 | 80.18 | 5.94 |
| cal51 | 1997 | 3 | 18 | 34.971 | -116.819 | 5.3 | 1.8 | 213 | 364 | 0.61 | 0.03 | **0.00** | - | 8.26 | 1.03 |
| cal52 * | 1997 | 4 | 26 | 34.369 | -118.670 | 5.1 | 1.6 | 301 | 364 | - | - | - | - | - | - |
| cal53 * | 1998 | 3 | 6 | 36.067 | -117.638 | 5.2 | 1.7 | 1127 | 362 | - | - | - | - | - | - |
| cal54 | 1999 | 10 | 16 | 34.594 | -116.271 | 7.1 | 3.6 | 150 | 335 | 1.15 | 0.05 | 0.03 | 0.01 | 17.53 | 2.23 |
| cal55 | 2001 | 2 | 10 | 34.289 | -116.946 | 5.1 | 1.6 | 865 | 363 | 0.36 | 0.02 | **0.00** | - | 12.64 | 1.29 |
| cal56 * | 2001 | 7 | 17 | 36.016 | -117.874 | 5.1 | 1.6 | 2229 | 363 | - | - | - | - | - | - |
| cal57 | 2001 | 10 | 31 | 33.508 | -116.514 | 5.1 | 1.6 | 441 | 364 | 0.66 | 0.02 | **0.00** | - | 20.62 | 1.61 |
| cal58 | 2002 | 2 | 22 | 32.319 | -115.322 | 5.7 | 2.2 | 480 | 362 | 0.97 | 0.03 | 0.11 | 0.04 | 55.33 | 5.65 |
| cal59 | 2003 | 2 | 22 | 34.310 | -116.848 | 5.4 | 1.9 | 305 | 358 | 0.82 | 0.02 | 0.00 | 0.00 | 21.21 | 1.63 |
| cal60 | 2003 | 12 | 22 | 35.709 | -121.104 | 6.5 | 3.0 | 130 | 340 | 0.91 | 0.05 | 0.00 | 0.02 | 11.96 | 1.90 |
| cal61 | 2004 | 9 | 29 | 35.390 | -118.624 | 5.0 | 1.5 | 132 | 92 | 0.80 | 0.03 | **0.00** | - | 12.18 | 1.16 |
| ita01 * | 1976 | 5 | 6 | 46.250 | 13.250 | 6.1 | 2.6 | 424 | 331 | - | - | - | - | - | - |
| ita02 | 1979 | 9 | 19 | 42.717 | 12.950 | 5.0 | 1.5 | 527 | 350 | 0.55 | 0.02 | **0.00** | - | 17.18 | 1.55 |
| ita03 | 1980 | 11 | 23 | 40.800 | 15.367 | 6.5 | 3.0 | 93 | 357 | 0.83 | 0.06 | 0.03 | 0.06 | 7.33 | 1.70 |
| ita04 | 1984 | 4 | 29 | 43.204 | 12.585 | 5.2 | 1.7 | 323 | 364 | 1.06 | 0.04 | 0.15 | 0.06 | 44.80 | 6.19 |
| ita05 * | 1984 | 5 | 7 | 41.666 | 13.820 | 5.8 | 2.3 | 351 | 357 | - | - | - | - | - | - |
| ita06 | 1990 | 5 | 5 | 40.650 | 15.882 | 5.6 | 2.1 | 129 | 364 | 0.96 | 0.04 | 0.02 | 0.02 | 12.31 | 1.76 |
| ita07 | 1995 | 9 | 30 | 41.790 | 15.971 | 5.4 | 1.9 | 131 | 343 | 0.84 | 0.06 | 0.12 | 0.13 | 11.25 | 2.76 |
| ita08 | 1996 | 10 | 15 | 44.799 | 10.679 | 5.5 | 2.0 | 177 | 362 | 1.05 | 0.06 | 0.25 | 0.13 | 26.91 | 5.64 |
| ita09 * | 1997 | 9 | 26 | 43.023 | 12.891 | 5.6 | 2.1 | 1592 | 345 | - | - | - | - | - | - |
| ita10 * | 1997 | 9 | 26 | 43.015 | 12.854 | 5.8 | 2.3 | 1070 | 345 | - | - | - | - | - | - |
| ita11 | 1998 | 9 | 9 | 40.060 | 15.949 | 5.6 | 2.1 | 277 | 361 | 0.37 | 0.04 | **0.00** | - | 4.17 | 0.75 |
| ita12 | 2002 | 9 | 6 | 38.381 | 13.654 | 5.6 | 2.1 | 688 | 358 | 1.48 | 0.08 | 3.68 | 0.89 | 695.33 | 253.84 |
| ita13 * | 2002 | 10 | 31 | 41.717 | 14.893 | 5.4 | 1.9 | 471 | 354 | - | - | - | - | - | - |
| ita14 | 2003 | 3 | 29 | 43.109 | 15.464 | 5.4 | 2.7 | 299 | 302 | 1.14 | 0.08 | 0.68 | 0.33 | 67.72 | 19.46 |
| ita15 | 2003 | 9 | 14 | 44.255 | 11.380 | 5.0 | 1.7 | 319 | 364 | 0.91 | 0.03 | 0.08 | 0.04 | 31.70 | 3.87 |

## Figure captions

Figure 1 – Effects on aftershock rate decay of parameter *c* of the modified Omori model (MOM). Panel a): aftershock rate as a function of time after the main shock for the MOM with $p$=1.2 and $c$ varying between 0 and 10 days. Panel b): behavior of effective $p$ (see text) as a function of time after the main shock for the same models (for $c$=0, effective $p$ coincide with $p$ at any time).

Figure 2 – Comparison among Omori-type models (eq. 1, blue, eq. 2, green, eq. 10 black, eq. 11, red) for Californian sequences. The numbers of sequences for which each model is the best are plotted as a function of the starting time $T_s$. In panel a) the comparison is based in $AIC_c$ while in panel b) on BIC.

Figure 3 – Comparison among Omori-type models as in Figure 1 but for Italian sequences.

Figure 4 – Comparison among Omori-type models as in Figure 1 but for the joint set of Californian and Italian sequences.

Figure 5 –Comparison, using BIC, of alternative models with Omori-type ones for the joint set of Californian and Italian sequences. STREXP with time shift (black) and without it (red) and LPL (green) are compared with the best Omori-type model (blue), in panel a), and with the MOM (blue), in panel b).

Figure 6 –Comparison between the MOM (solid) and the model of eq. (2) with $c$ fixed to 0 (dotted), as a function of starting time $T_s$, for the joint set of Californian and Italian sequences.

Figure 7 –Comparison between MOM (solid) and the model of eq. (2) with $c$ fixed to 0 (dotted), as a function of the difference between minimum magnitude of aftershocks and main shock magnitude, for Californian sequences.

Figure 8 –Comparison, as in Figure 7, between MOM (solid) and the model of eq. (2) with $c$ fixed to 0 (dotted) but for Italian sequences.

Figure 9 – Behavior of aftershock decay model parameters as a function of starting time $T_s$ for all 47 analyzed sequences. In panel a) the parameter $p$ of eq. (2), in panels b) and c) the parameters of the MOM (eq. 11) $p$ and $c$ respectively. In panels a) and b) the thick dashed lines indicate the arithmetic average among all sequences of parameter $p$, while in panel c) the geometric average of non-null $c$ values.

Figure 10 – Same as Fig. 9 but as a function of the difference between minimum and main shock magnitudes.

**Figure 1**

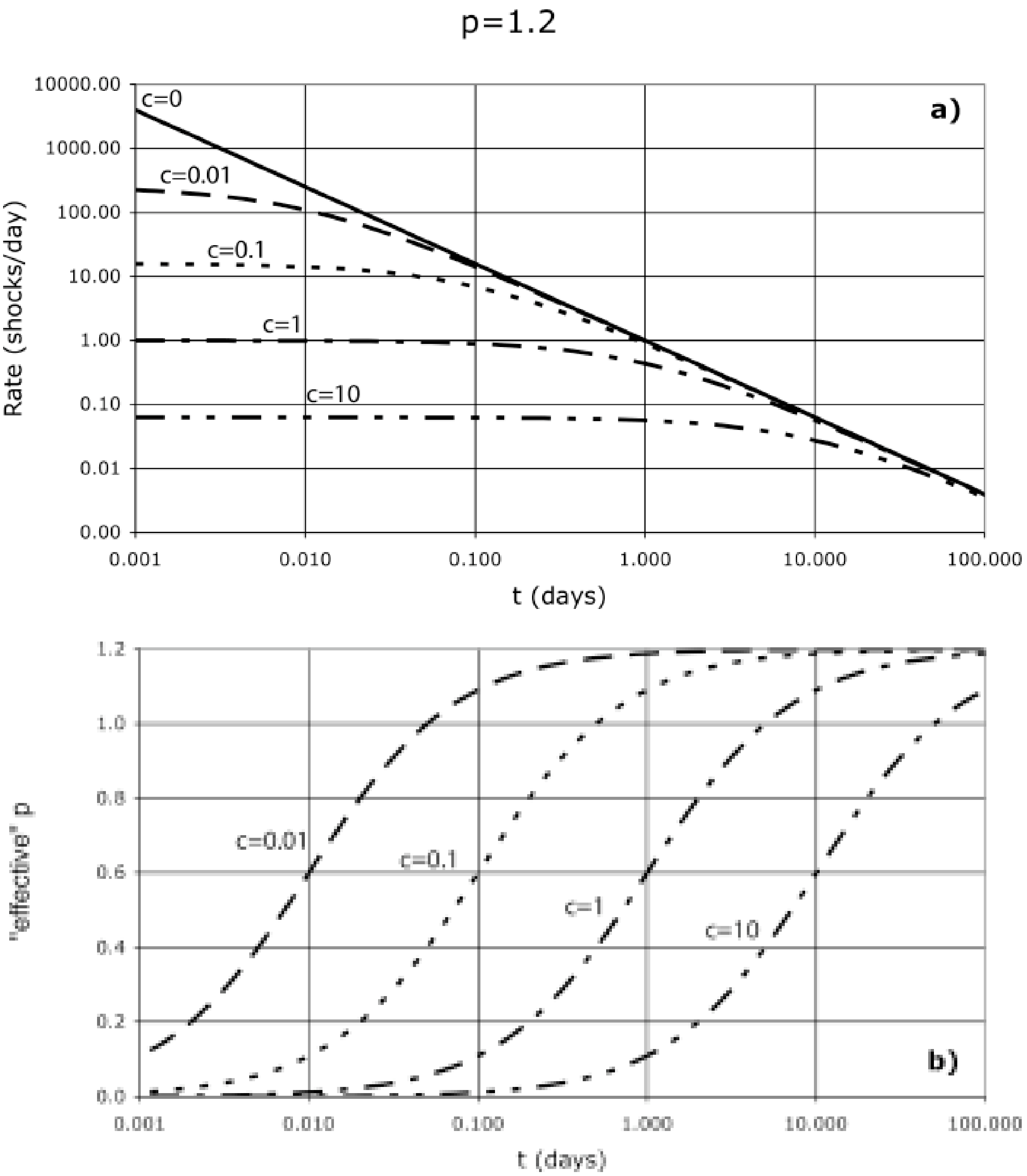

# Figure 2

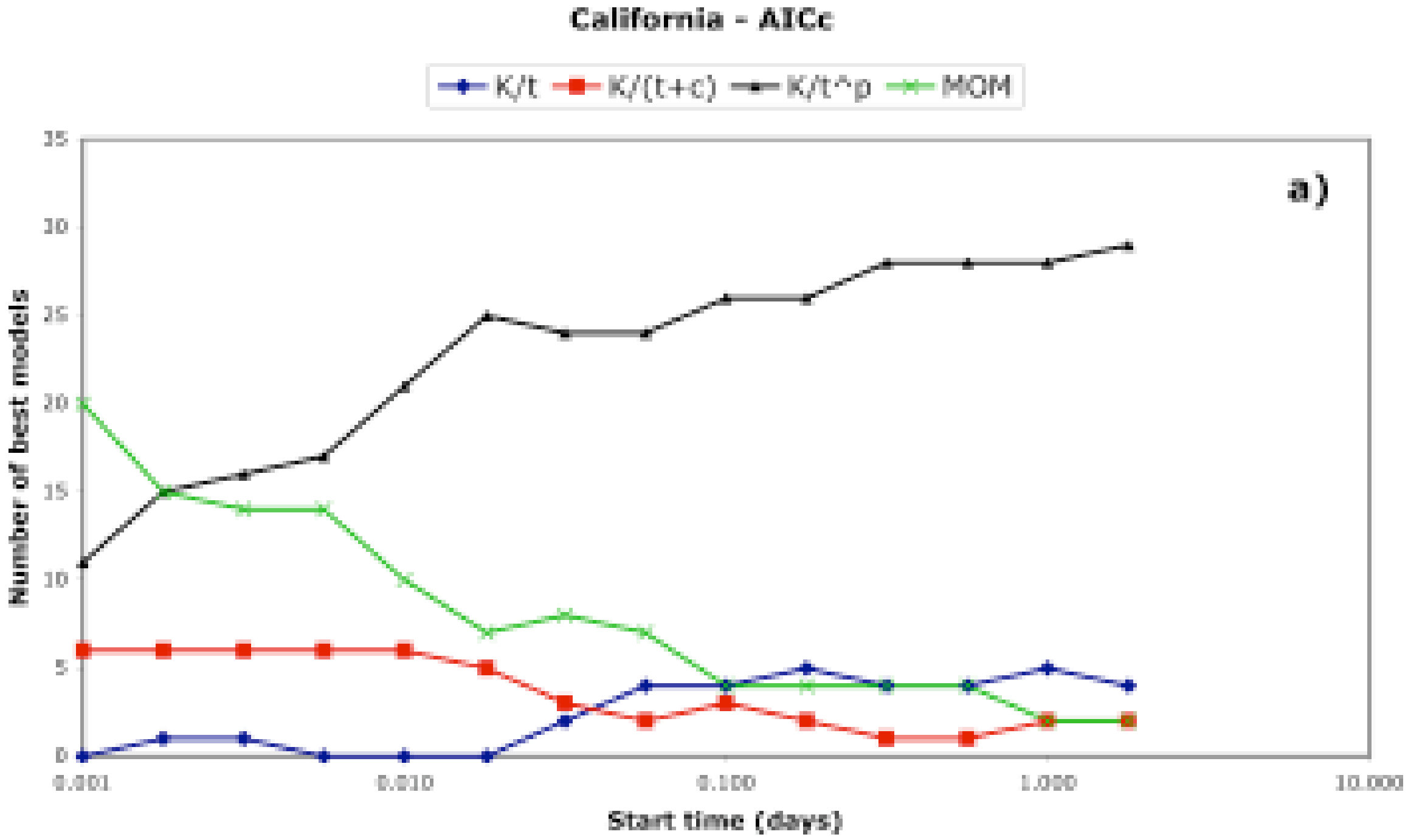

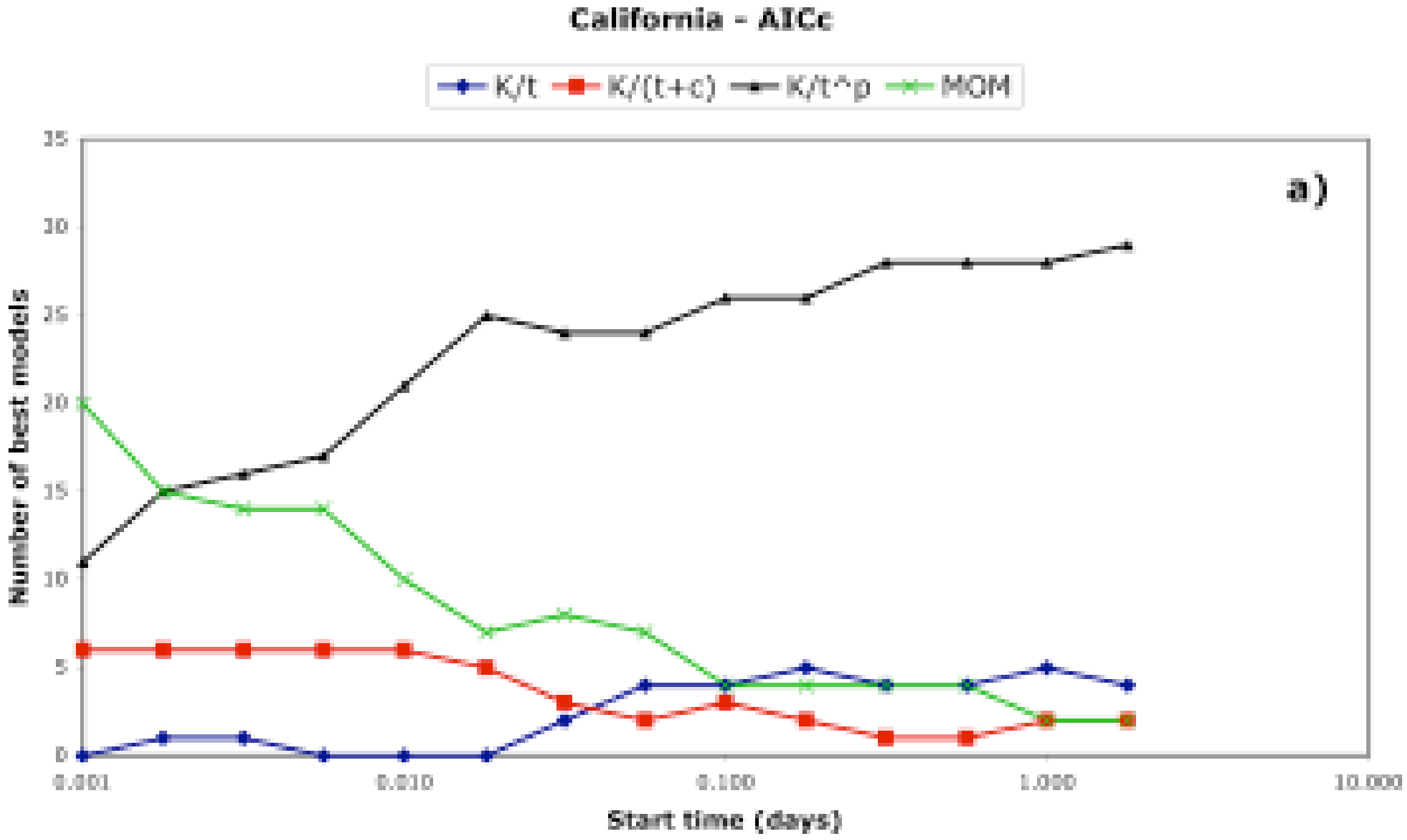

# Figure 3

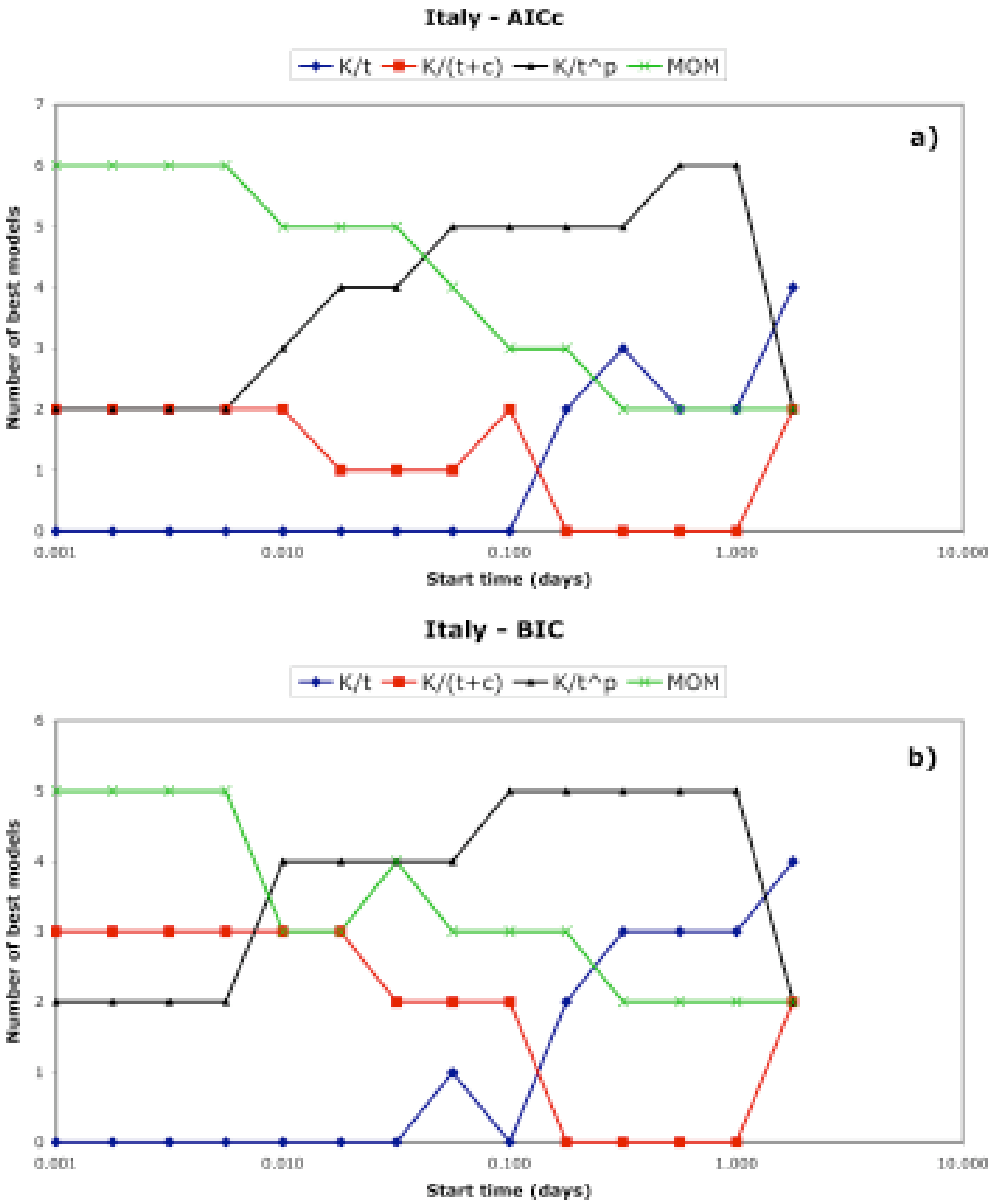

**Figure 4**

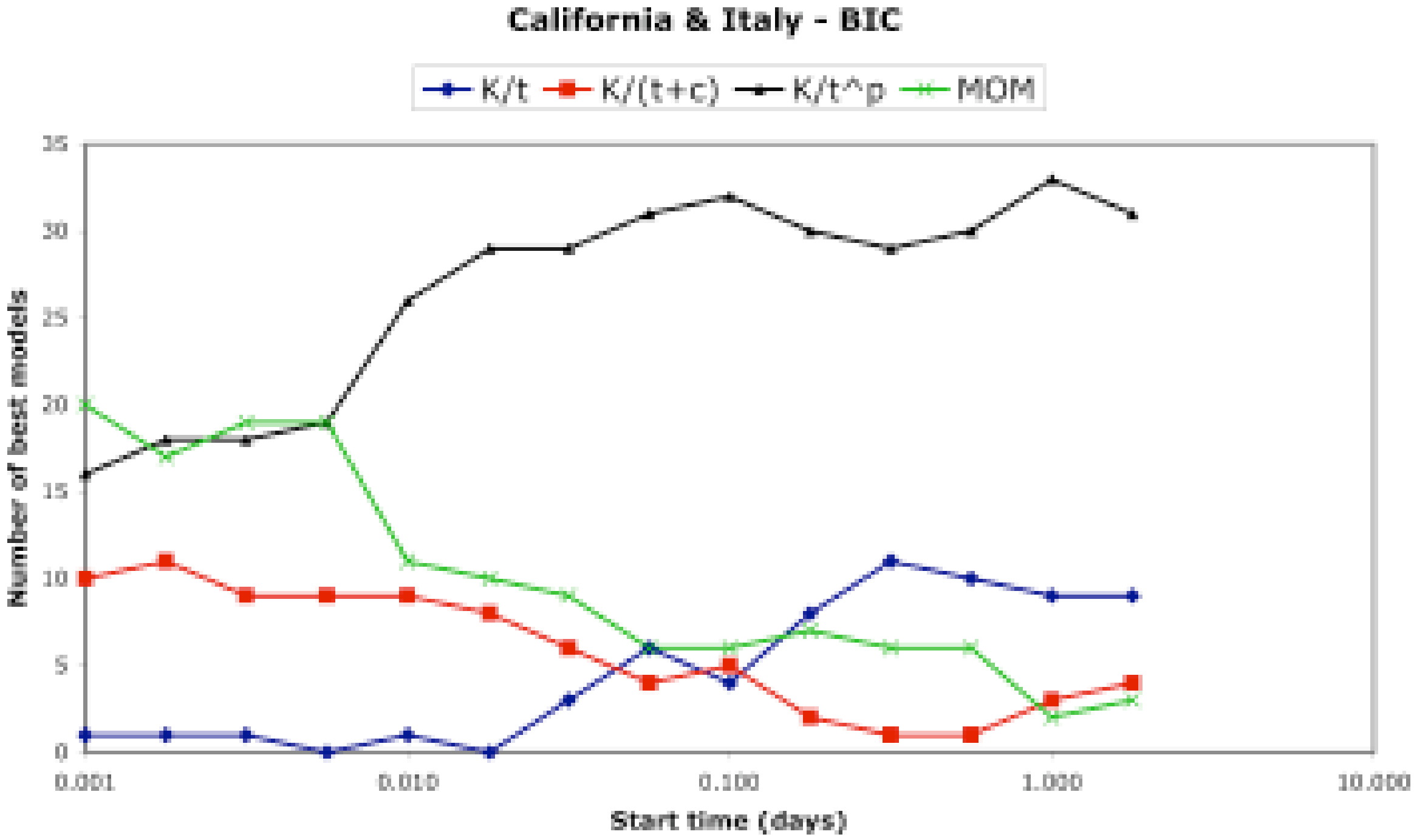

**Figure 5**

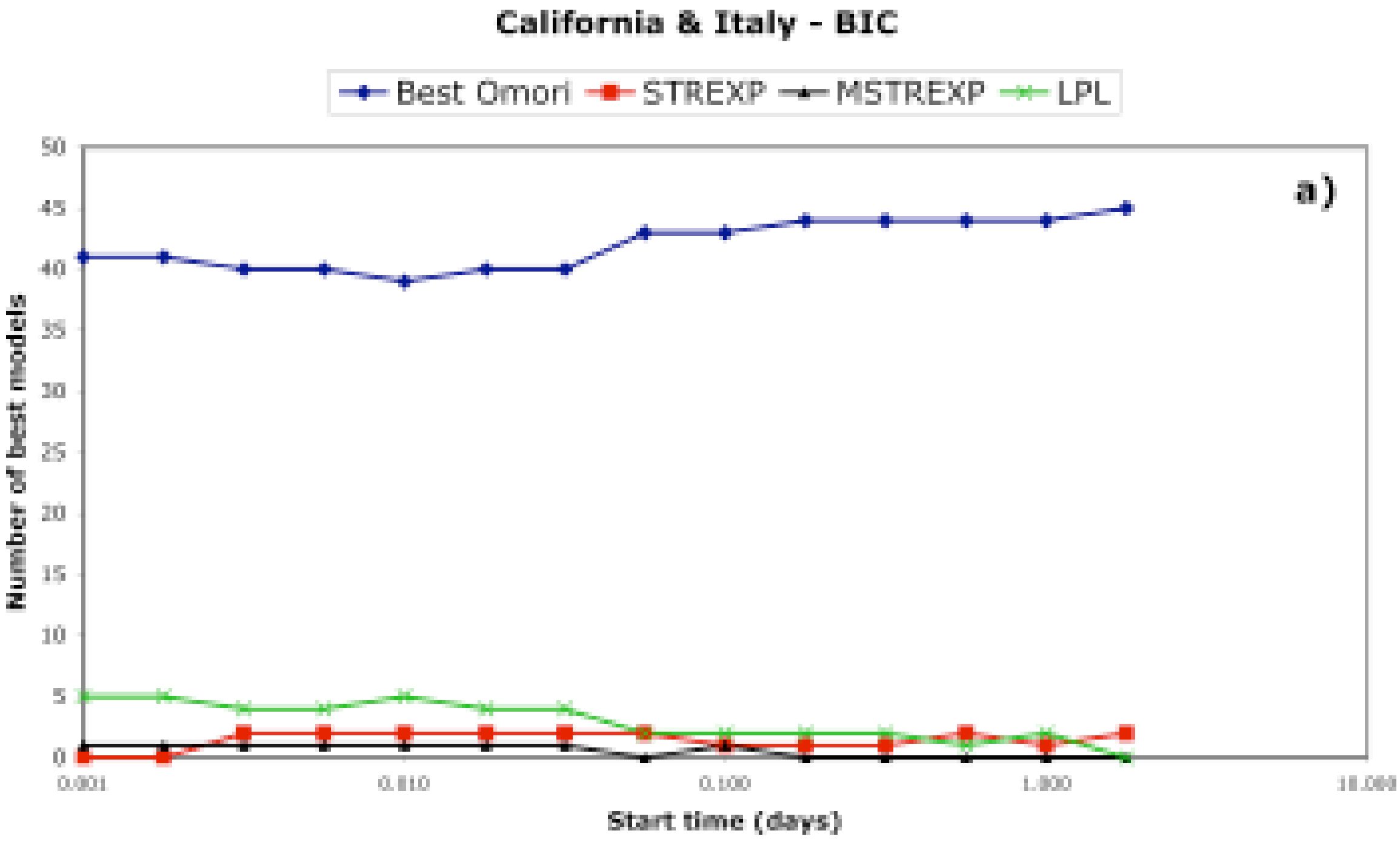

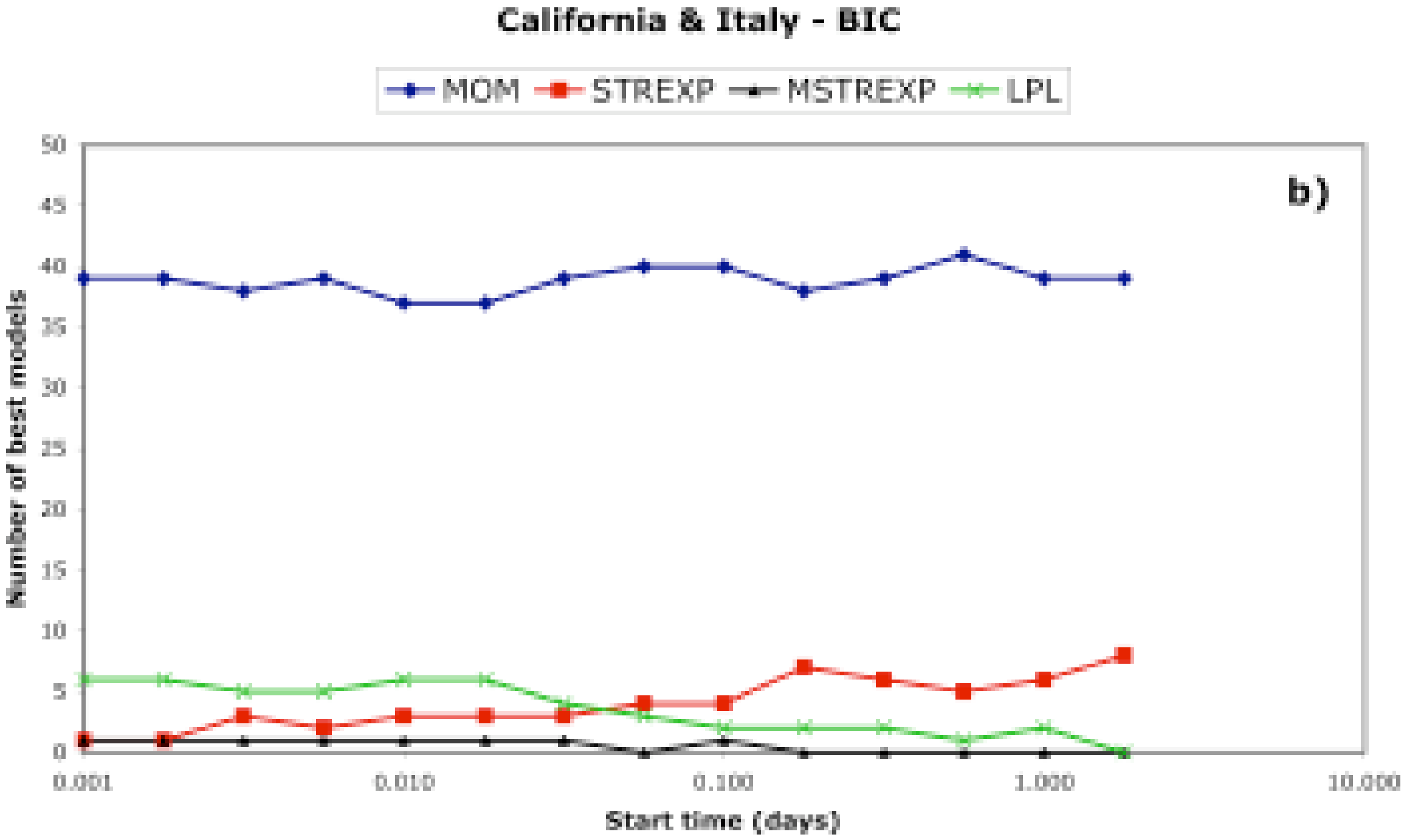

**Figure 6**

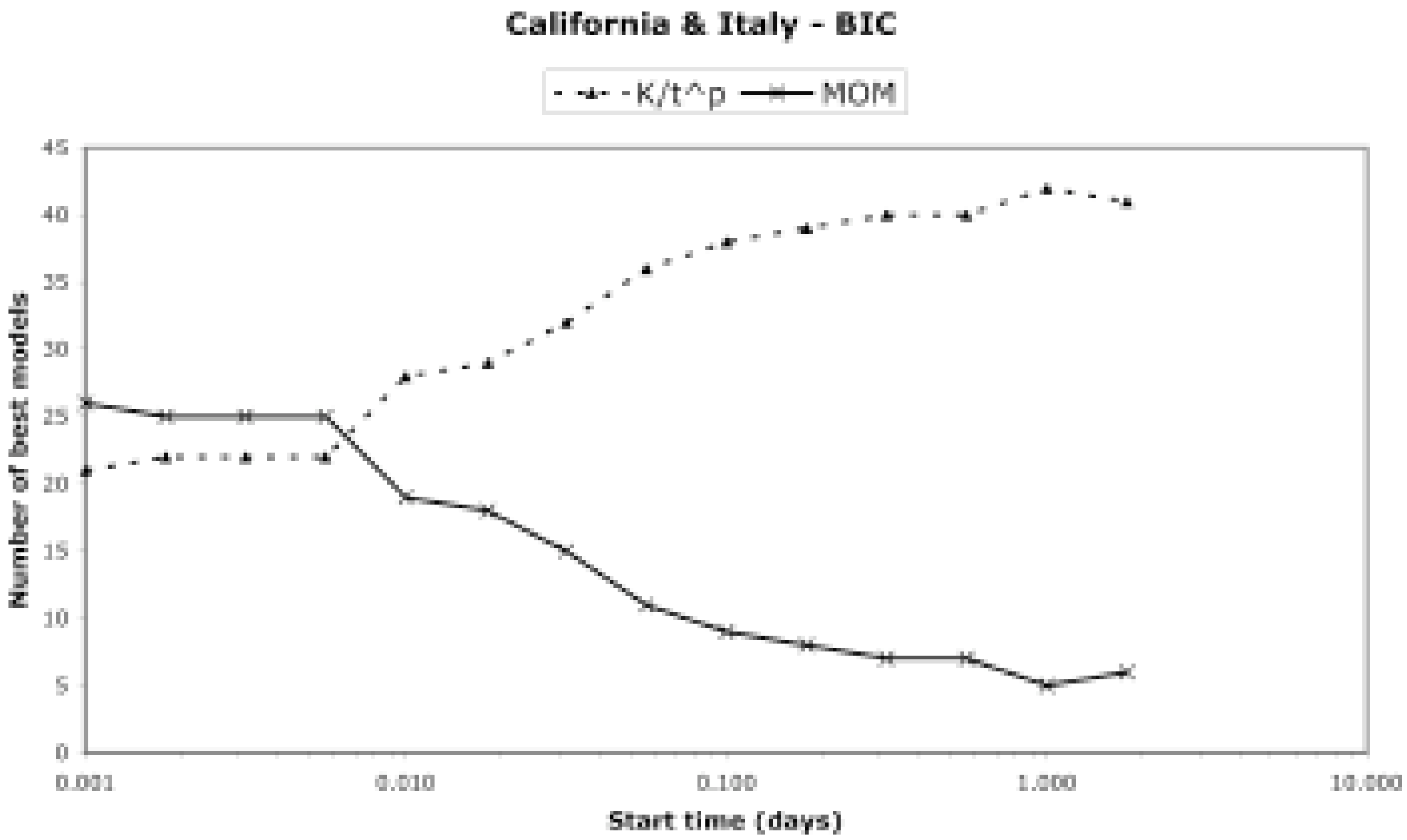

**Figure 7**

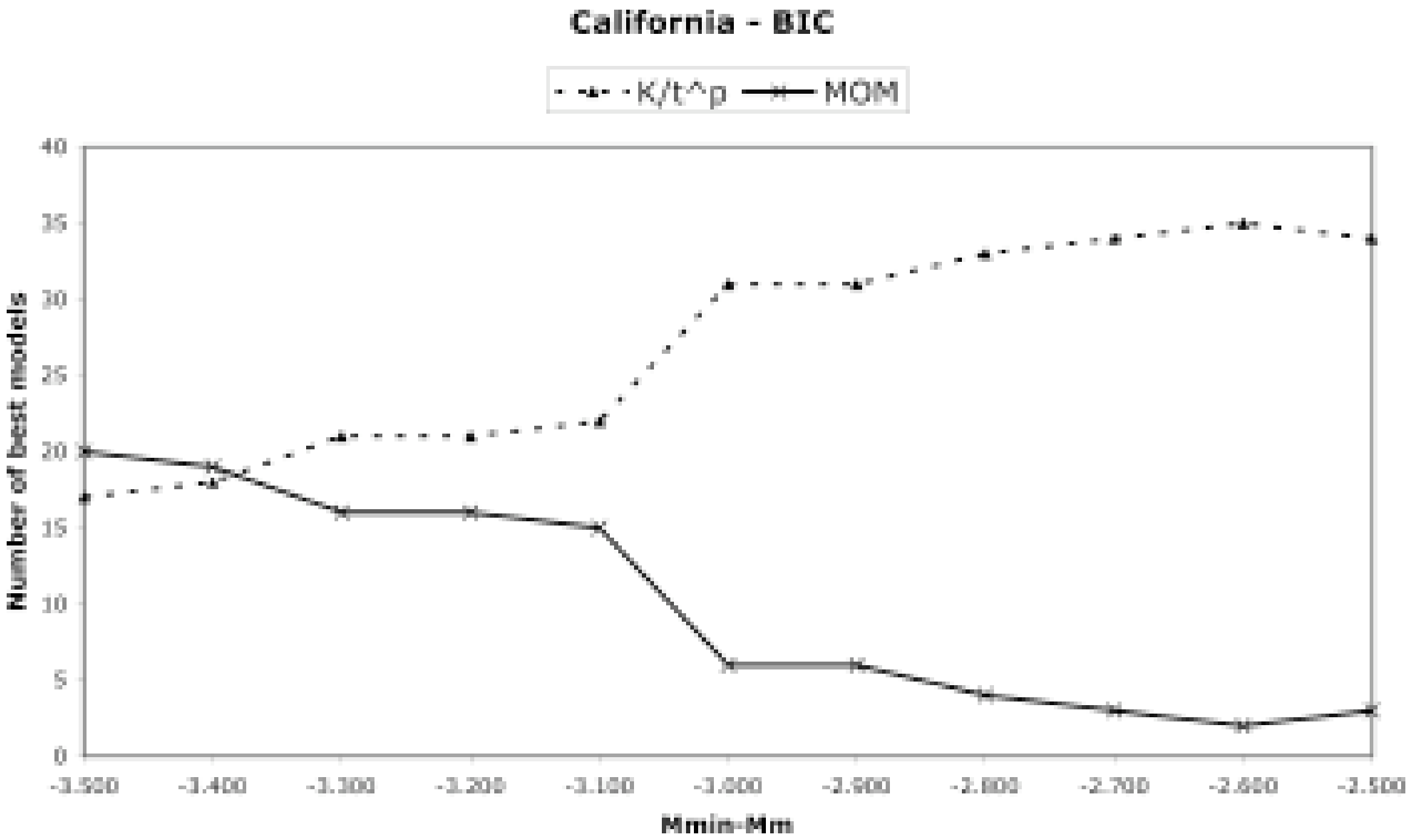

**Figure 8**

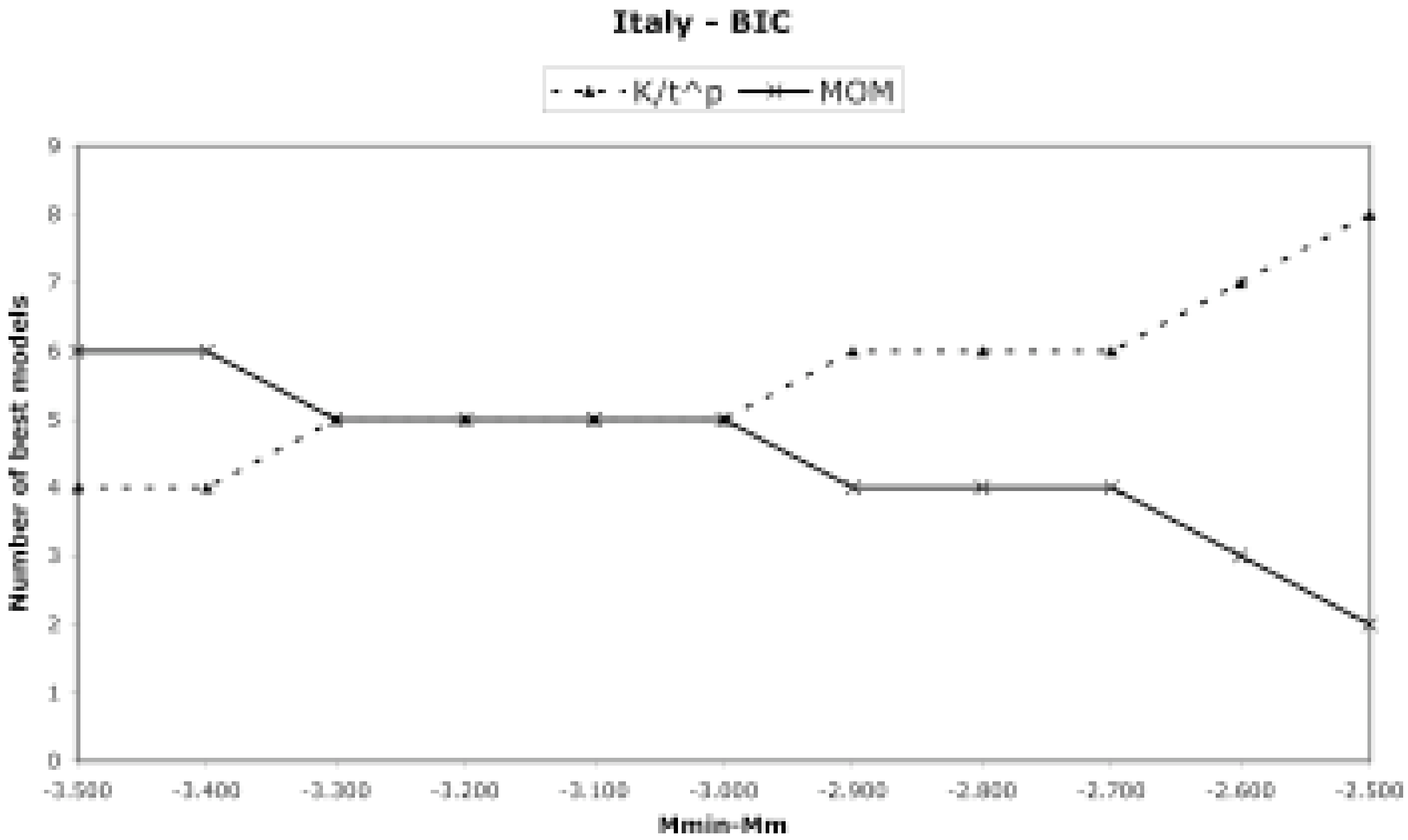

# Figure 9

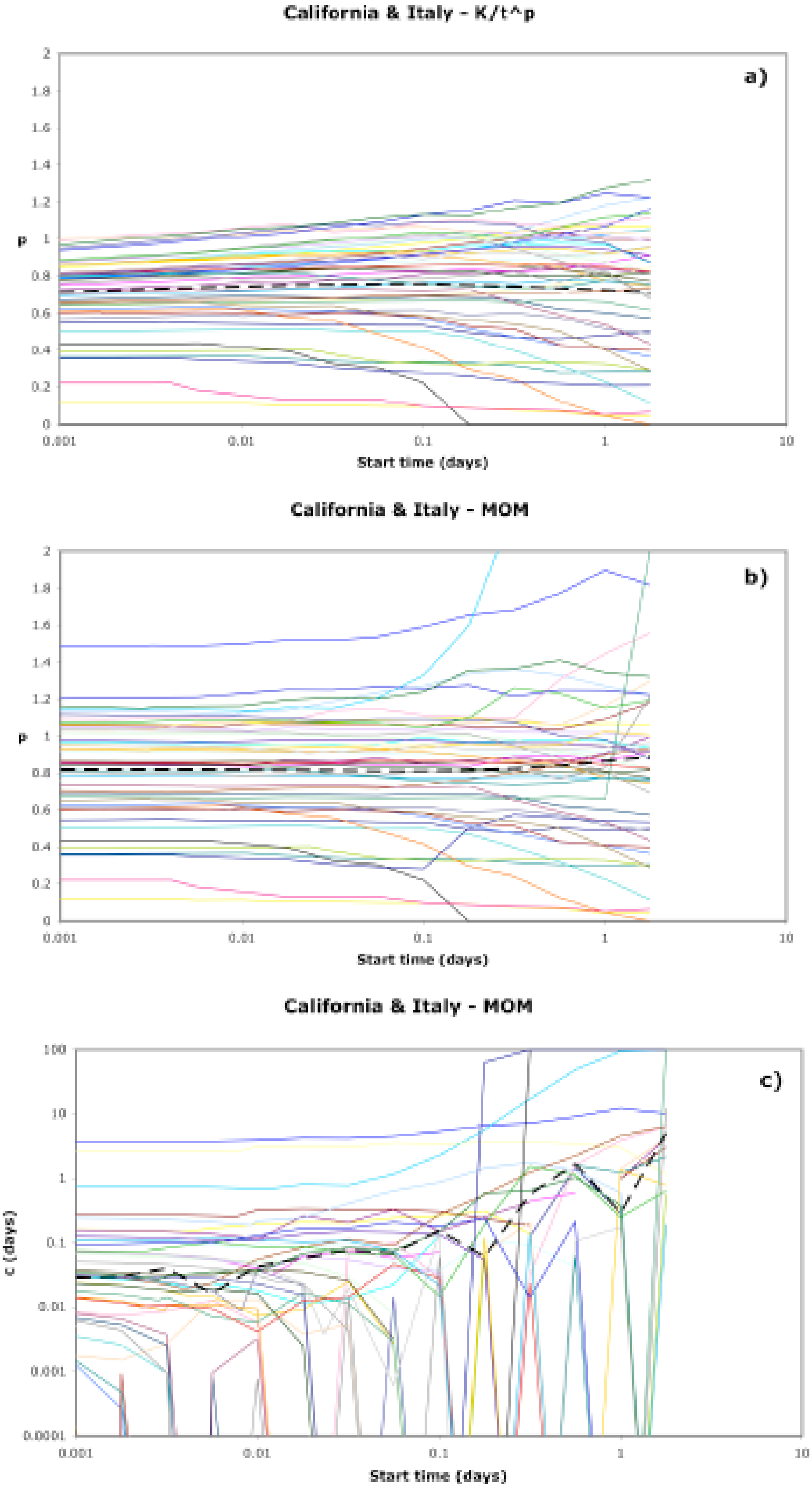
California & Italy - K/t^p
a)
p
Start time (days)
California & Italy - MOM
b)
p
Start time (days)
California & Italy - MOM
c)
c (days)
Start time (days)

# Figure 10

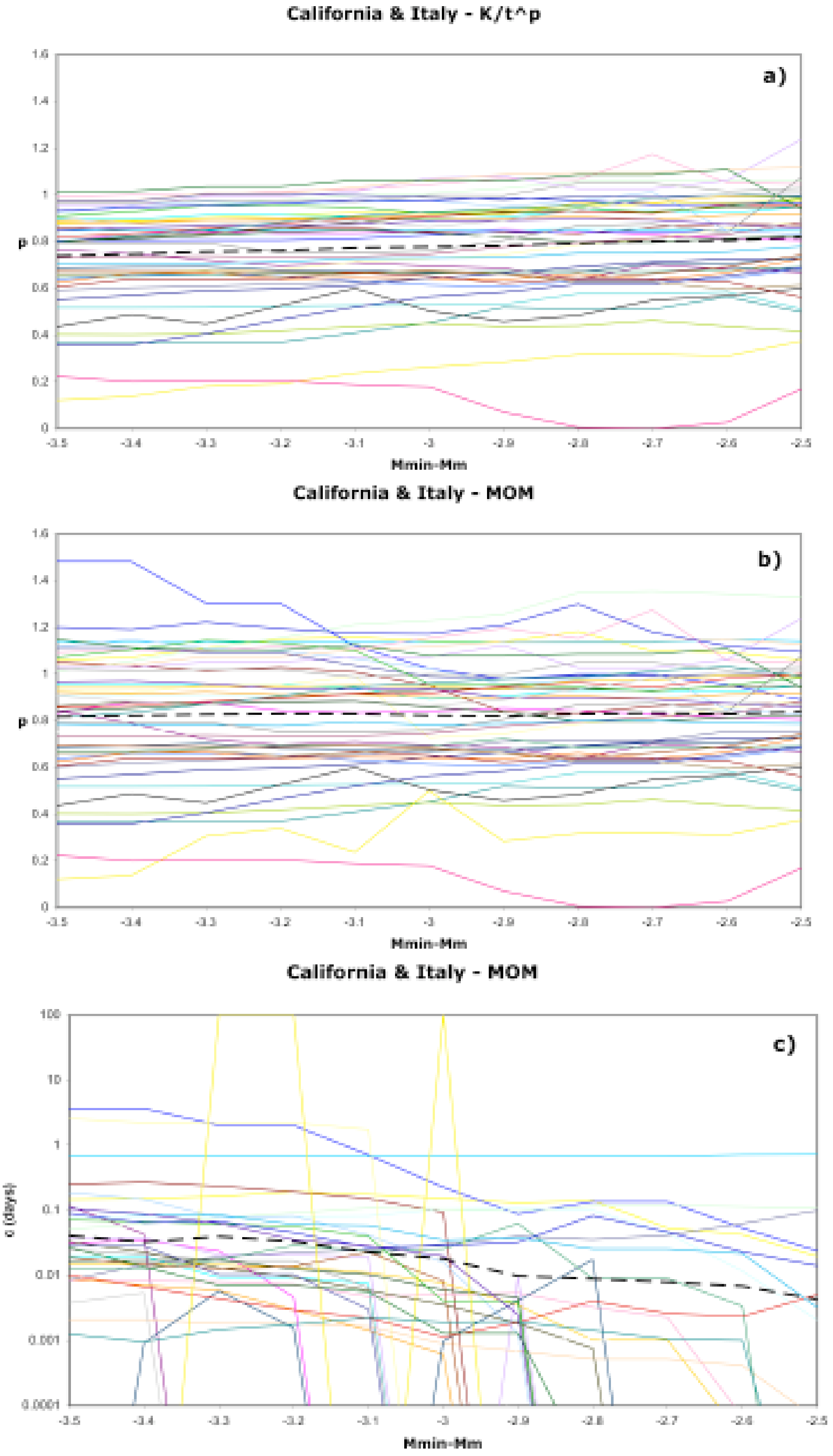